\documentclass[floatfix,reprint,superscriptaddress,amsmath,amssymb,pra,noeprint]{revtex4-1}
\pdfoutput=1

\usepackage[left=1.5cm, right=1.5cm, top=1.785cm, bottom=2.0cm]{geometry}
\usepackage{graphicx}
\usepackage{setspace}
\usepackage[usenames, dvipsnames]{color}
\usepackage{braket}
\usepackage{upgreek}
\usepackage{ulem}
\usepackage[english]{babel}
\usepackage{dcolumn}
\usepackage{bm}

\newcommand{\figref}[1]{fig.~\ref{#1}}

\begin{document}

\title{Coherent Manipulation of the Internal State of Ultracold \textsuperscript{87}Rb\textsuperscript{133}Cs Molecules with Multiple Microwave Fields}

\author{Jacob A. Blackmore}
\email[email: ]{j.a.blackmore@durham.ac.uk}
\affiliation{Joint Quantum Centre (JQC) Durham-Newcastle, Department of Physics, Durham University, South Road, Durham, DH1 3LE}

\author{Philip D. Gregory}
\affiliation{Joint Quantum Centre (JQC) Durham-Newcastle, Department of Physics, Durham University, South Road, Durham, DH1 3LE}

\author{Sarah L. Bromley}
\affiliation{Joint Quantum Centre (JQC) Durham-Newcastle, Department of Physics, Durham University, South Road, Durham, DH1 3LE}

\author{Simon L. Cornish}
\email[email: ]{s.l.cornish@durham.ac.uk}
\affiliation{Joint Quantum Centre (JQC) Durham-Newcastle, Department of Physics, Durham University, South Road, Durham, DH1 3LE}

\begin{abstract}
We explore coherent multi-photon processes in \textsuperscript{87}Rb\textsuperscript{133}Cs molecules using 3-level lambda and ladder configurations of rotational and hyperfine states, and discuss their relevance to future applications in quantum computation and quantum simulation. In the lambda configuration, we demonstrate the driving of population between two hyperfine levels of the rotational ground state via a two-photon Raman transition. Such pairs of states may be used in the future as a quantum memory, and we measure a Ramsey coherence time for a superposition of these states of 58(9)\,ms. In the ladder configuration, we show that we can generate and coherently populate microwave dressed states via the observation of an Autler-Townes doublet. We demonstrate that we can control the strength of this dressing by varying the intensity of the microwave coupling field. Finally, we perform spectroscopy of the rotational states of \textsuperscript{87}Rb\textsuperscript{133}Cs up to $N=6$, highlighting the potential of ultracold molecules for quantum simulation in synthetic dimensions. By fitting the measured transition frequencies we determine a new value of the centrifugal distortion coefficient $D_v=h\times207.3(2)$\,Hz. 

\end{abstract} 
\maketitle
Coherent control of complex quantum systems is of great importance for the development of new quantum technologies. Ultracold polar molecules are one such system which has attracted much interest, motivated by the combination of rich internal molecular structure and the accessibility of strong dipole-dipole interactions. These properties have led to many proposals for using ultracold polar molecules for quantum computation~\cite{Demille2002,Yelin2006,Zhu2013,Herrera2014,Ni2018,Sawant2020,Hughes2020}, quantum simulation~\cite{Barnett2006,Micheli2006,Buchler2007,Macia2012,Manmana2013,Gorshkov2013}, quantum-state controlled chemistry~\cite{Krems2008,Bell2009,Ospelkaus2010,Dulieu2011,Balakrishnan2016,Hu2019}, and precision measurement of fundamental constants~\cite{Zelevinsky2008,Hudson2011,Salumbides2011,Salumbides2013,Schiller2014,ACME2014,Hanneke2016,Cairncross2017,Borkowski2018,ACME2018,Borkowski2019}. A number of experiments have successfully generated trapped gases of polar molecules at ultracold temperatures either by association of pre-cooled atomic gases~\cite{Ni2008,Takekoshi2014,Molony2014,Park2015,Guo2016,Rvachov2017,Seesselberg2018,Hu2019,Yang2019,Voges2020} or by direct laser cooling~\cite{Shuman2010,Hummon2013,Zhelyazkova2014,Barry2014,McCarron2015,Norrgard2016,Kozyryev2017,Truppe2017,Lim2018,Anderegg2018}. 
Most recently, the former method was used to create the first Fermi-degenerate gas of ultracold polar molecules~\cite{DeMarco2019}. 

The vast majority of the proposed applications of ultracold molecules utilise the rotational and hyperfine degrees of freedom, which together form a large and rich internal space. Using a pair of rotational states connected via a non-zero transition dipole moment leads to effective spin-exchange interactions~\cite{Yan2013}, opening up applications in the simulation of quantum magnetism~\cite{Barnett2006,Micheli2006,Gorshkov2011,Gorshkov2011b,Zhou2011,Manmana2013,Hazzard2013}. Long-range interactions between molecules can also be engineered by preparing superpositions of rotational states, using either microwave or DC electric fields. 
Under such conditions, molecules confined in an optical lattice where tunneling between sites is possible are predicted to exhibit a range of novel quantum phases~\cite{Buchler2007,Micheli2007,Pollet2010,Capogrosso_Sansone2010,Macia2012,Lechner2013,Gorshkov2013}. 
It is also possible to use hyperfine levels of the rotational ground state of polar molecules as a quantum memory. Here the molecules do not interact via the dipole-dipole interaction and long coherence times for superpositions of these states are possible~\cite{Park2017}. Coupling such states to strongly-interacting levels in the molecule enables quantum-gate operations and applications in quantum computation. 
Moreover, by introducing large numbers of hyperfine or rotational states we can use the molecules as multi-level qudits, greatly expanding the computational space and improving scalability~\cite{Sawant2020}. Furthermore, it has recently been proposed that the rotational states of polar molecules may be used to engineer fully controllable synthetic dimensions to experiments~\cite{Sundar2018,Sundar2019}. With these applications in mind, a number of groups, including our own, continue to develop the necessary techniques to coherently control and fully exploit the internal degrees of freedom of molecules~\cite{Ospelkaus2010,Will2016,Gregory2016,Guo2018,Gong2019,Ji2020}.

In this paper, we present multi-photon coherent control of the hyperfine and rotational states of ultracold \textsuperscript{87}Rb\textsuperscript{133}Cs molecules (hereafter RbCs). We first study a three-level lambda-type configuration, and show that by detuning both driving fields from resonance we can drive Raman transitions between hyperfine levels of the rotational ground state. Using Ramsey interferometry, we confirm the generation of coherent superpositions of these states through the observation of high-contrast fringes. We then consider a three-level ladder-type system; by setting a strong driving field on resonance, and using a second weak field as a probe, we observe an Autler-Townes doublet indicating the controlled production of coherent dressed states. In search of a larger Hilbert space, we demonstrate through a sequence of microwave transfers that we can coherently populate rotationally excited states up to $N=6$. Taken together, these developments lay the foundations for the use of RbCs for quantum simulations and illustrate the potential of ultracold molecules as a platform for new quantum technologies.

\section{Theory}\label{sec:Theory}
We begin with a brief overview of the relevant theory to describe the rotational and hyperfine structure of $^1\Sigma$ diatomic molecule in the vibronic ground state. In an external magnetic field, this structure is described by a Hamiltonian comprised of three terms~\cite{Brown&Carrington,Aldegunde2008}:
\begin{equation}
	H_\mathrm{RbCs} = H_\mathrm{rot}+H_\mathrm{hf}+H_\mathrm{Zeeman}.
	\label{eq:Hamiltonian}
\end{equation}
Here $H_\mathrm{rot}$ describes the rotational structure, $H_\mathrm{hf}$ describes the hyperfine structure and $H_\textrm{Zeeman}$ describes the interaction between the molecule and an external magnetic field. We can write these terms explicitly~\cite{Herzberg,Brown&Carrington,Aldegunde2008, Aldegunde2017}:
\begin{subequations}\label{eq:Hyperfine}
	\begin{align}
		\label{eqn:Rotational}
		H_\mathrm{rot} &= B_v \bm{N}^2 -D_v \bm{N}^2\cdot\bm{N}^2,\\ 
		\label{eqn:Hyperfine}
		\begin{split}
			H_\mathrm{hf} &= \sum_{j=\mathrm{Rb}, \mathrm{Cs}} e\bm{Q}_{j} \cdot \bm{q}_{j}-c_{3}\sqrt{6}\bm{T}^2(C)\cdot \bm{T}^2\left(\bm{I}_\mathrm{Cs},\bm{I}_\mathrm{Rb}\right) \\
			&+ c_{4} \bm{I}_{\mathrm{Rb}} \cdot \bm{I}_{\mathrm{Cs}}
			+\sum_{j=\mathrm{Rb}, \mathrm{Cs}} c_{j} \bm{N} \cdot \bm{I}_{j},
		\end{split}\\
		\label{eqn:Zeeman}
		H_\mathrm{Zeeman} &=-g_\mathrm{r} \mu_\mathrm{N} \bm{N} \cdot \bm{B}-\sum_{j=\mathrm{Rb}, \mathrm{Cs}} g_{j}\left(1-\sigma_{j}\right) \mu_{\mathrm{N}} \bm{I}_{j} \cdot \bm{B}.
	\end{align}
\end{subequations}
The rotational contribution $H_{\text{rot}}$ \eqref{eqn:Rotational} is defined by the rotational angular momentum operator~$\bm{N}$, and the rotational and centrifugal distortion constants, $B_{v}$ and $D_{v}$. The hyperfine contribution $H_{\text{hf}}$ \eqref{eqn:Hyperfine} consists of four terms. The first describes the electric quadrupole interaction and represents the interaction between the nuclear electric quadrupole of nucleus~$j$ ($e\bm{Q}_j$) and the electric field gradient at the nucleus ($\bm{q}_j$). The second and third terms represent the tensor and scalar interactions between the nuclear magnetic moments, with tensor and scalar spin-spin coupling constants $c_{3}$ and $c_{4}$ respectively, and $\bm{I}_\mathrm{Rb}$, $\bm{I}_\mathrm{Cs}$ are the vectors for the nuclear spin of \textsuperscript{87}Rb and \textsuperscript{133}Cs respectively. The fourth term is the interaction between the nuclear magnetic moments and the magnetic field generated by the rotation of the molecule, with spin-rotation coupling constants $c_{\text{Rb}}$ and $c_{\text{Cs}}$. Finally, the Zeeman contribution $H_{\text{Zeeman}}$ \eqref{eqn:Zeeman} consists of two terms which represent the rotational and nuclear interaction with an externally applied magnetic field. The rotation of the molecule produces a magnetic moment which is characterized by the rotational $g$-factor of the molecule ($g_{r}$). The nuclear interaction similarly depends on the nuclear $g$-factors ($g_{\text{Rb}}$,~$g_{\text{Cs}}$) and nuclear shielding ($\sigma_{\text{Rb}}$,~$\sigma_{\text{Cs}}$) for each species. We use the constants tabulated by Gregory~\textit{et al.}\cite{Gregory2016} when calculating the energy levels and eigenstates of $H_{\rm{RbCs}}$ \eqref{eq:Hamiltonian}.

At zero magnetic field, the quantum states of RbCs are well described by the quantum numbers $(N,F)$.
Here $N$ is the rotational quantum number and the associated energy is $B_v N (N+1)$, which leads to splittings between neighbouring rotational states in the microwave domain (for RbCs, $B_v\approx490$\,MHz). $F$ is the resultant from the addition of the rotational angular momentum and the nuclear spins ($I_\mathrm{Rb}=3/2$ and $I_\mathrm{Cs}=7/2$).
In the ground rotational state, $N=0$, there are four values of $F$: 2, 3, 4, and 5. Applying a magnetic field splits each state $F$ into separate Zeeman sublevels labelled by $M_F$, the projection of $F$ along the space-fixed axis defined by the magnetic field (from now on we label this the $z$-axis). In RbCs, this results in each rotational state $N$ being comprised of $32\times(2N+1)$ hyperfine Zeeman sub-levels. In the limit of large magnetic fields, the rotational and nuclear angular momenta decouple and the hyperfine Zeeman sub-levels become uniquely identified by the quantum numbers $(N, M_N, m_\mathrm{Rb}, m_\mathrm{Cs})$, where $M_N, m_\mathrm{Rb}, m_\mathrm{Cs}$ are the projections of the rotational angular momentum of the molecule and the nuclear spins, respectively. The experiments we present in this work take place at a magnetic field of 181.5\,G. This field is not high enough to decouple the rotational and nuclear angular momenta, and the only good quantum numbers available are $N$ and $M_F$. As this is not sufficient to uniquely identify a given hyperfine state, we label hyperfine states in the molecule by $(N, M_F)_k$ where $k$ is an index counting up the states in order of increasing energy, such that $k=0$ is the lowest energy state for given values of $N$ and $M_F$.

Transitions between rotational states can be driven by microwave fields with resolved hyperfine sub-levels. The transitions that are electric dipole allowed are those where $\Delta N =1$ and $\Delta M_F=0,\pm1$. The strength of the transition is determined by the transition dipole moment 
\begin{equation}
	\bm{\mu_{i,j}} = \bra{\psi_i}\bm{\mu}\ket{\psi_j},
\end{equation}
where the components ($\mu^z_{i,j}$,$\mu^+_{i,j}$,$\mu^-_{i,j}$) of $\boldsymbol{\mu}_{i,j}$ describe the strength of $\pi$, $\sigma^+$ and $\sigma^-$ transitions respectively. 

\section{Experimental Apparatus}\label{sec:Apparatus}
Our experimental apparatus and methodology has been discussed in detail elsewhere~\cite{McCarron2011, Koeppinger2014, Molony2014, Gregory2015, Molony2016,Molony2016a} and so only a short summary is given here. We produce RbCs molecules from an ultracold mixture of \textsuperscript{87}Rb and \textsuperscript{133}Cs atoms using magnetoassociation on an interspecies Feshbach resonance at 197\,G~\cite{Koehler2006,Chin2010,Koeppinger2014}. Evaporative cooling of the atomic gases and the association take place in a crossed optical dipole trap, which operates at a wavelength $\lambda=1550~\mathrm{nm}$. The optical dipole trap light is linearly polarised such that the polarisation is parallel to the applied magnetic field. The molecules are transferred to the $M_F=5$ hyperfine sub-level of the $X^1\Sigma\,(v=0,\,N=0)$ rovibrational ground state  using stimulated Raman adiabatic passage (STIRAP)~\cite{Bergmann1998,Molony2014,Gregory2015,Molony2016a}. To avoid spatially-varying AC Stark shifts the STIRAP is performed in free-space~\cite{Molony2016a}. At our operating magnetic field of 181.5~G, the $M_F=5$ hyperfine sub-level is the lowest in energy and is therefore the absolute ground state of the molecule. We detect molecules by reversing the creation process and imaging the resulting atomic clouds. As such, only molecules that undergo a second STIRAP sequence are imaged. Because the STIRAP process is state-selective we can only image molecules in the $X^1\Sigma\,(v=0,\,N=0,\,M_F=5)$ hyperfine sub-level. 

We coherently drive transitions between pairs of rotational states with controlled microwave pulses\cite{Gregory2016}. The lowest frequency microwaves we require are for the transition between the ground and first-excited rotational states. These microwaves have a frequency of $2\times B_v\approx980$\,MHz. To drive this transition we use a pair of homebuilt quarter-wavelength monopole antennas oriented perpendicularly to one another \textit{i.e.} one is oriented along $z$ and the other in the $x{\text -}y$ plane. In free space these antennas would emit microwaves linearly polarised along their length, however finite-element method modelling (using COMSOL Multiphysics RF) indicates that when placed into our apparatus they both emit a significant $z$-polarised component due to the boundary conditions imposed by the surrounding magnetic field coils. Access to higher rotational states requires higher frequency microwaves, for which we use a broadband microwave horn (Atlantec-rf AS-series). The microwaves propagate from the horn towards the molecules at an angle of approximately 30$^\circ$ from $z$, such that it can drive $\pi$, $\sigma^+$ and $\sigma^-$ transitions. The microwave signals are generated by commercial signal generators (a pair of Keysight MXG N5183B and an Agilent E8257D), which are synchronised to a common 10\,MHz GPS reference (Jackson Labs Fury) and connected to 3\,W amplifiers. We define the power input to the antennas as the power output by the amplifiers. Pulses are generated using either the built-in pulse modulation mode on the signal generators or an external switch, and are controlled by transistor-transistor logic (TTL) signals derived from a field programmable gate array (FPGA) with microsecond timing resolution.

\section{Raman Transitions in a 3-level Lambda System}\label{sec:Raman}
\begin{figure}[t!]
	\centering
	\includegraphics[width=0.45\textwidth]{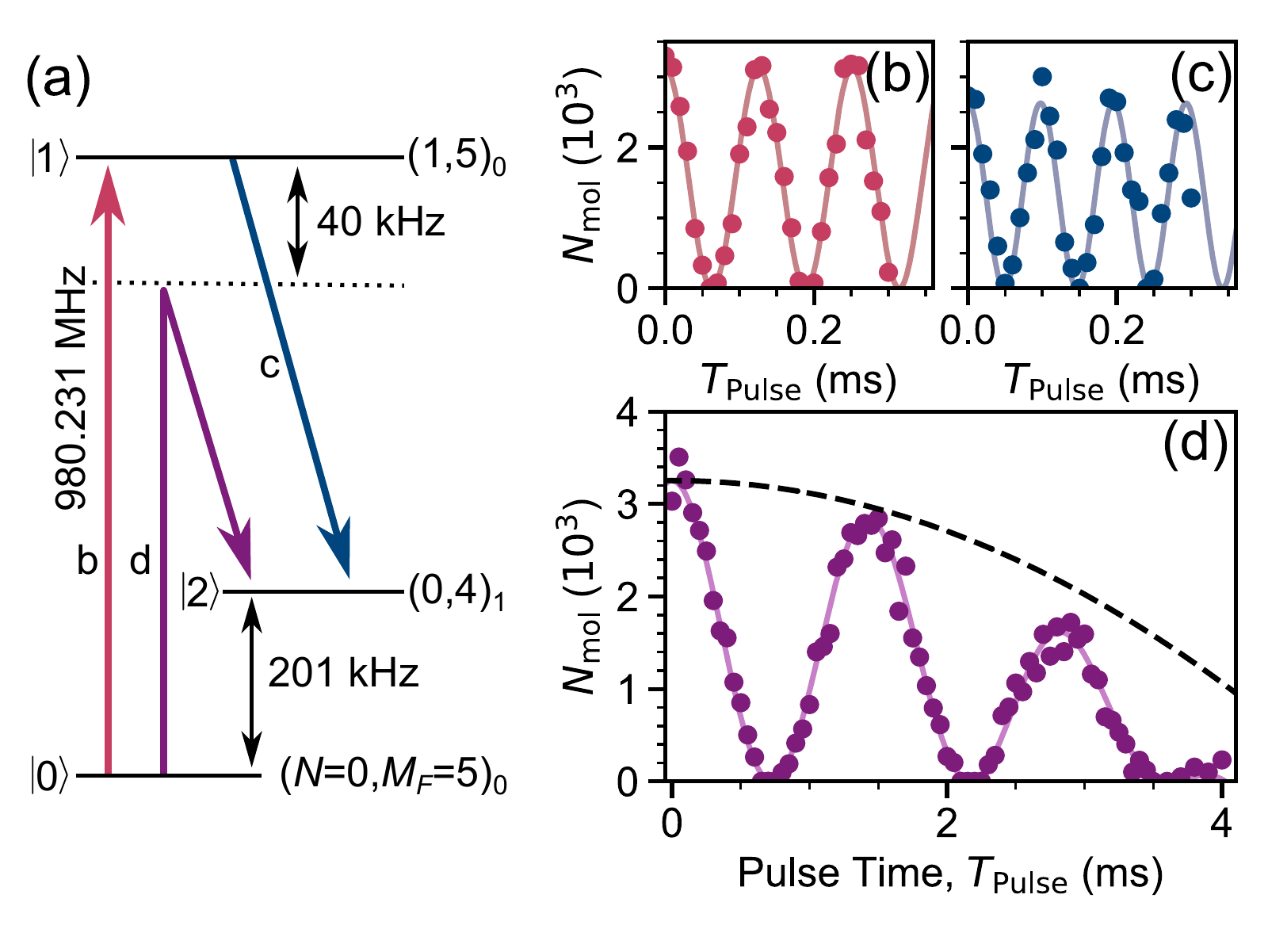}
	\caption{One- and two-photon Rabi oscillations in the 3-level lambda system. (a)~The energy level configuration used. The coloured lines b, c and d correspond to the transitions shown in (b-d). The dotted line indicates the $\Delta/2\pi = 40~\mathrm{kHz}$ detuning used for the Raman transition. (b)~Resonant one-photon Rabi oscillations on the $\ket{0}\rightarrow\ket{1}$ transition. (c)~Resonant one-photon Rabi oscillations on the $\ket{1}\rightarrow\ket{2}$ transition. (d)~Two-photon Rabi oscillations on the Raman transition between $\ket{0}$ and $\ket{2}$. The loss of molecules with increasing pulse time is due to the untrapped molecules leaving the the detection region. An independent measurement of the loss of molecules is indicated by the dashed line.
	}
	\label{fig:Rabi_oscillations}
\end{figure}

We begin our investigations by studying a 3-level lambda configuration of states, consisting of two hyperfine levels of the rotational ground state coupled to a common rotationally excited state. The initial ground state $\ket{0}=\ket{N=0, M_F=5}_0$ is fixed by the STIRAP, and to form the lambda system, we choose $\ket{1}=\ket{1,5}_0$ and $\ket{2}=\ket{0,4}_1$ as depicted in \figref{fig:Rabi_oscillations}(a). The state $\ket{1}$ is chosen as it is the lowest energy hyperfine level of $N=1$, which simplifies the scheme and enables detuning of the microwaves without the risk of off-resonant excitation of other transitions. The state $\ket{2}$ is subsequently chosen to maximise the relevant coupling to the rotationally excited state $\bm{\mu_{1,2}}$. The full state compositions of $\ket{0},\ket{1}$ and $\ket{2}$, in the uncoupled basis, are given in Appendix \ref{App:States}

This type of system is promising for the initialisation of a quantum memory, where the two hyperfine levels of the rotational ground state $\ket{0},\ket{2}$ define the stored qubit. For a quantum memory to be effective, long-lived coherence is required. These states are well suited for this, as molecules stored in these states experience the same polarisability, leading to the possibility of long coherence times in optical traps\cite{Gregory2017,Park2017,Blackmore2020}. In contrast, for qubits constructed from two \emph{different} rotational states, differential ac Stark shifts arising from the anisotropy of the polarisability are typically the primary cause of decoherence for optically trapped molecules\cite{Neyenhuis2012,Seesselberg2018_Coherence,Blackmore2018}.

We first demonstrate one-photon coherent control in this system. In \figref{fig:Rabi_oscillations}(b) we show the one-photon Rabi oscillations resulting from driving the $\ket{0}\rightarrow\ket{1}$ transition. The microwaves are set on resonance with the transition, with frequency \mbox{$f=980.231$\,MHz}, and we measure a Rabi frequency \mbox{$\Omega_{0,1}/2\pi = 7.95(3)~\mathrm{kHz}$}. To drive the $\ket{1}\rightarrow\ket{2}$ transition, we first transfer the population to $\ket{1}$ using a $\pi$-pulse on $\ket{0}\rightarrow\ket{1}$. We then pulse on the microwaves resonant with $\ket{1}\rightarrow\ket{2}$, $f=980.039$\,MHz, for a variable time before returning the population back to $\ket{0}$ for detection using a second $\pi$-pulse on $\ket{0}\rightarrow\ket{1}$. The resultant Rabi oscillations are shown in \figref{fig:Rabi_oscillations}(c), and we measure a Rabi frequency $\Omega_{1,2}/2\pi = 10.23(7)~\mathrm{kHz}$.

To properly initialise the quantum memory it is important that the states can be prepared without populating $\ket{1}$, as partially populating states in $N=1$ introduces an additional mechanism for dephasing due to the differential polarisability. In order to achieve direct coupling between $\ket{0}$ and $\ket{2}$ we employ a two-photon Raman transition. To drive the Raman transition between $\ket{0}\rightarrow\ket{2}$, we introduce a one-photon detuning of $\Delta = 2\pi \times 40(2)~\mathrm{kHz} \gg \Omega_{0,1}, \Omega_{1,2}$, but remain on two-photon resonance. By pulsing both microwave fields on simultaneously, with Rabi frequencies as in \figref{fig:Rabi_oscillations}(b-c), we observe two-photon Rabi oscillations with an effective Rabi frequency of $\Omega^{'}_{0,2}/2\pi = 691.4(1.8)~\mathrm{Hz}$ as shown in \figref{fig:Rabi_oscillations}(d). 
As the spectroscopy is performed in free-space, the sample expands due to its thermal energy and falls due to gravity. This leads to a perceived loss as a function of time as molecules exit the imaging volume defined by the waists (about $30$\,$\upmu$m) of the STIRAP beams. The dashed line shown in \figref{fig:Rabi_oscillations}(d) indicates the number of molecules remaining in the detection region, as measured independently with no microwaves present. The Rabi oscillations we observe are highly coherent, with no significant loss of contrast during the available interrogation time. The two-photon effective Rabi frequency we can achieve is currently limited by the available microwave power and the need to avoid coupling to other nearby states in the molecule.

\begin{figure}[t]
	\centering
	\includegraphics[width=0.45\textwidth]{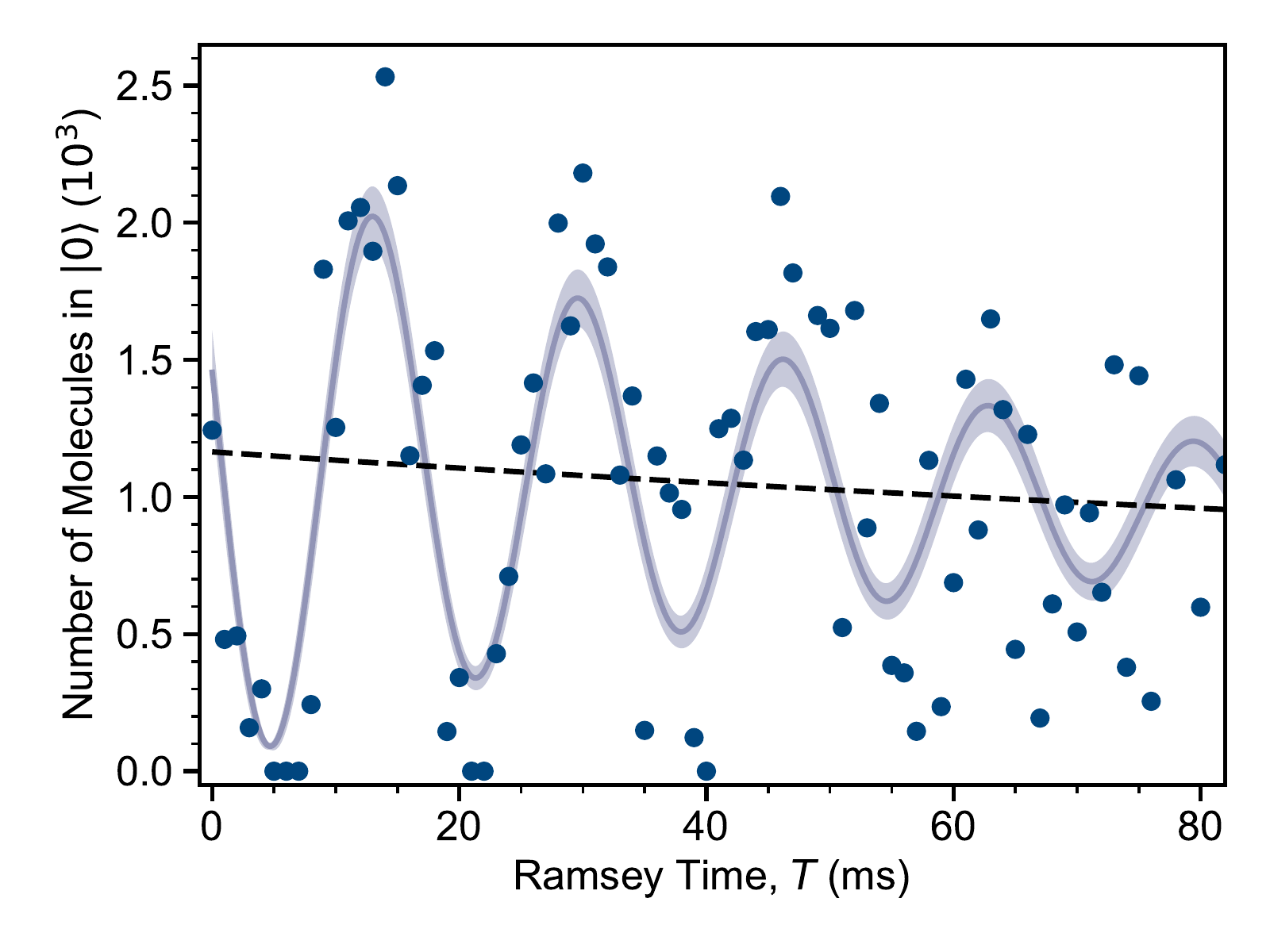}
	\caption{Ramsey fringes observed between $\ket{0}=\ket{0,5}_0$ and $\ket{2}=\ket{0,4}_1$, in the 3-level lambda system shown in \figref{fig:Rabi_oscillations}(a). A two-photon Raman $\pi/2$-pulse of duration $\mathrm{361.5~\upmu s}$ is used to generate a superposition of $\ket{0}$ and $\ket{2}$ in free-space. The molecules are recaptured in the optical dipole trap where we allow the superposition to evolve for a time $T$. Following this time, a second $\pi/2$-pulse, identical to the first, is used to project the phase relative to the microwave field onto the population in $\ket{0}$. The Ramsey fringes decay with a 1/$e$ time of $T_2=58(9)$\,ms. The total number of molecules reduces over the course of the measurement with a $1/e$ time $T_1=0.48(11)$~s due to optical excitation of two-body complexes formed during  molecular collisions~\cite{Gregory2020}. The dashed line indicates the associated expected decay in the number of molecules in $\ket{0}$ in the absence of any coherence. The 60.1(6)~Hz oscillation frequency of the fringes shown here indicates that the microwave fields are not two-photon resonant, and enables high-precision measurement of the energy of the hyperfine sub-levels.
	}
	\label{fig:Ramsey_fringes}
\end{figure}

To test the coherence, we create a superposition of $\ket{0}$ and $\ket{2}$ by using a $\pi/2$ pulse with duration $361.5~\upmu\mathrm{s}$. The molecules are then recaptured by turning on the optical dipole trap for a time $T$, during this time the superposition is allowed to evolve freely. Following the hold, we perform another $\pi/2$ pulse. This projects the phase of the prepared superposition onto the populations of the states, which we read out by measuring the number of molecules $N(T)$ in state $\ket{0}$. The resulting Ramsey fringes are shown in \figref{fig:Ramsey_fringes}. 

We fit a model to our data which accounts for both collisional loss of molecules and dephasing of the superposition,
\begin{equation}
	N(T) = N_i \left(\frac{1}{1+\frac{T}{T_1}\times[e-1]}\right)\times\frac{1}{2}\times\left[e^{-T/T_2}\cos(\delta T +\phi)+1\right].
\end{equation}
Here, $N_i$ is the initial total number of molecules, $T_1$ is the $1/e$ lifetime for molecules in the trap, $T_2$ is the 1/$e$ dephasing time, and $\delta$ and $\phi$ are the frequency and phase of the Ramsey fringes. Our measured value of $T_1$ is 0.48(11)~s, this is consistent with a lifetime limited by fast optical excitation of long-lived two-body collision complexes~\cite{Gregory2019,Gregory2020}. 
The fringes we observe indicate a two-photon detuning of our microwaves of $\delta/2\pi = 60.1(6)~\mathrm{Hz}$. The small uncertainty in this measurement of the detuning shows how this technique can be used to measure the relative energies between hyperfine states to sub-Hz precision. Combined with the known frequencies of our microwaves, we measure an energy difference between $\ket{0}$ and $\ket{2}$ of $h\times201.043\,8(6)~\mathrm{kHz}$, which is consistent with the theoretical prediction of $h\times201.3(1.2)~\mathrm{kHz}$. 
The superposition decoheres with a $1/e$ time of $T_2=58(9)$~ms. This is two orders of magnitude longer than the longest coherence time we have previously measured for superpositions of different rotational states~\cite{Blackmore2018}. 

A differential Zeeman shift between $\ket{0}$ and $\ket{2}$ is likely the primary cause for decoherence in our experiment. To quantify this source of decoherence we calculate the differential magnetic moment
\begin{equation}
	\Delta\mu_\textrm{mag}=\bra{0}\mu_z\ket{0}-\bra{2}\mu_z\ket{2},
\end{equation} 
where   
\begin{equation}
	\mu_z =-g_\mathrm{r} \mu_\mathrm{N} \bm{N} \cdot \hat{z}-\sum_{j=\mathrm{Rb}, \mathrm{Cs}} g_{j}\left(1-\sigma_{j}\right) \mu_{\mathrm{N}} \bm{I}_{j} \cdot \hat{z},
\end{equation}
is the component of the magnetic moment that lies along the magnetic field.
For our chosen states the differential magnetic moment is \mbox{$h\times1.3~\mathrm{kHz\,G^{-1}}$}. We estimate that the magnetic field stability in our experiment is about $10$\,mG. This translates to a frequency stability for the transition of $13$\,Hz, or a coherence time of 77\,ms, in agreement with our observations. Park~\textit{et al.} have studied a similar configuration of states in \textsuperscript{23}Na\textsuperscript{40}K molecules\cite{Park2017}. In their system, the differential Zeeman shift is an order of magnitude smaller than in our experiments and they observe a correspondingly longer coherence time. 
To increase the coherence time of the superposition between two hyperfine sub-levels in RbCs we should therefore look for a pair of states that have the smallest possible differential magnetic moment. At 181.5~G our calculations indicate that a superposition of $(0,4)_1$ and $(0,3)_0$ has a differential magnetic moment of $h\times 71~\mathrm{Hz\,G^{-1}}$ which, with our current level of magnetic field stability, would lead to a coherence time of 250~ms.

\section{Autler-Townes in a 3-level Ladder System}\label{sec:AT}

\begin{figure}[t]
	\centering
	\includegraphics[width=0.45\textwidth]{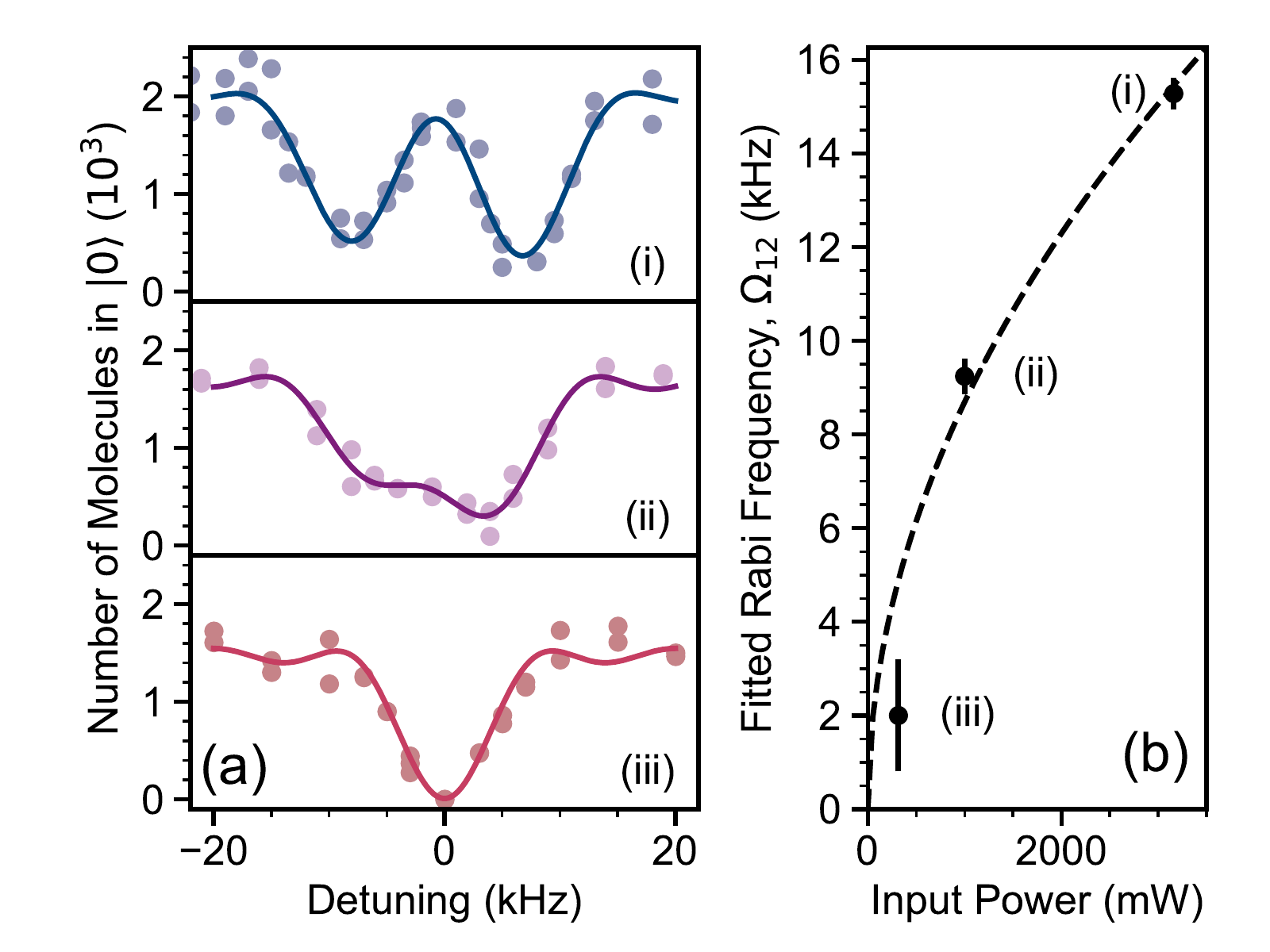}
	\caption{Autler-Townes spectroscopy in the 3-level ladder system, $(N,M_F)_i=(0,5)_0\leftrightarrow(1,6)_0\leftrightarrow(2,7)_0$.
		(a) The number of molecules remaining in $(0,5)_0$, while varying the frequency of the probe field weakly coupling the $(0,5)_0 \rightarrow (1,6)_0$ transition ($\Omega_{01}/2\pi = 5.2(1)~\mathrm{kHz}$). Three measurements are shown, labelled (i) to (iii), demonstrating the effect of decreasing the strength of the $(1,6)_0\leftrightarrow(2,7)_0$ coupling field. The solid lines are fits to \eqref{eq:schrodingereq}, with $\Omega_{12}$ and $\Delta_{12}$ as free parameters. The fitting reveals that the asymmetry observed in the spectra is explained by a coupling-field detuning of $\Delta_{12}/2\pi =-1.7(5)~\mathrm{kHz}$. (b) The Rabi frequencies $\Omega_{12}$, as a function of the input power to the microwave horn. 
	}
	\label{fig:Autler-Townes}
\end{figure}

We now consider a 3-level ladder configuration of states and demonstrate the generation of coherent dressed states via the observation of an Autler-Townes doublet. This configuration has several applications. Gorshkov~\textit{et~al.}\cite{Gorshkov2011, Gorshkov2011b} have shown theoretically that pairs of coherently dressed states in molecules can be used as spin-states to realise a wide range of highly tunable $t$-$J$-$V$-$W$ models, featuring long-range spin-spin interactions $J_z$ and $J_\perp$ of $XXZ$ type, long-range density-density interactions $V$, and long-range density-spin interactions $W$. The interactions in these models are controlled by tuning a combination of the microwave dressing fields and a DC electric field. In addition, it has been shown that microwave dressing can be used to modify the rate of collisions between pairs of molecules, causing resonant alignment of molecules as they approach each other during a collision, and leading to strong attractive forces\cite{Yan2020}. It has been predicted that microwave dressing can also be used to suppress collisional losses, by using circularly-polarised microwaves to engineer repulsive long-range interactions between the molecules to prevent pairs of molecules reaching short-range\cite{Lassabliere2018,Karman2018}. 

We construct the ladder using the spin-stretched states \mbox{$\ket{0}=\ket{0,5}_0$}, \mbox{$\ket{1}=\ket{1,6}_0$} and \mbox{$\ket{2} = \ket{2,7}_0$}. As previously, the molecules are initialised in $\ket{0}$ by the STIRAP. We dress the initially unpopulated state $\ket{1}$ with a component of $\ket{2}$ using microwaves near-resonant with the $\ket{1}\rightarrow\ket{2}$ transition. To probe the dressed state, we use spectroscopy on the $\ket{0}\rightarrow\ket{1}$ transition. In \figref{fig:Autler-Townes}(a) we show spectroscopy of the dressed rotational state for increasing power on the $\ket{1}\rightarrow\ket{2}$ transition. 

We model the interaction of the molecule with the microwave field using a simplified three-level Hamiltonian
\begin{equation}\label{eq:3level}
	\hat{H}_\textrm{3-level} =\frac{\hbar}{2} \begin{pmatrix} 
		0 & \Omega_{01} & 0\\
		\Omega_{01} & -2\Delta_{01} & \Omega_{12}\\
		0 & \Omega_{12} & -2(\Delta_{12}-\Delta_{01})\\
	\end{pmatrix},
\end{equation}
where $\Omega_{jk}$ and $\Delta_{jk}$ are the Rabi frequency and detuning of the field driving the $\ket{j}\rightarrow\ket{k}$ transition. We measure $\Omega_{01}/2\pi= 5.2(1)~\textrm{kHz}$ by direct observation of Rabi oscillations. We fit our results for the population remaining in $\ket{0}$ as a function of $\Delta_{01}$ with a numerical solution to the Schr\"odinger equation
\begin{equation}
	\frac{\mathrm{d}}{\mathrm{d}t}\ket{\psi} =-\frac{i}{\hbar} \hat{H}_\textrm{3-level}\ket{\psi},
	\label{eq:schrodingereq}
\end{equation}
where $\Omega_{12}$ and $\Delta_{12}$ are free parameters in the fitting. 

As the power is increased we observe a clear splitting between the two dressed states that increases with the square-root of power, shown in \figref{fig:Autler-Townes}(b), as is expected for an Autler-Townes doublet. The maximum amount of power we can supply to the microwave horn is limited by the 3~W amplifiers. At maximum power, we observe a splitting of $\Omega_{12}=15.3(3)$~kHz. We also note that there is a slight asymmetry in the observed lines. This is caused by a small detuning, $\Delta_{12} = -1.7(5)$~kHz, of the coupling field. 

We can interpret the results presented here in the context of a quantum simulation of a single particle in a 1D lattice consisting of only two sites. In this language the states $\ket{1}$ and $\ket{2}$ represent the sites of the lattice, represented by the synthetic dimension constructed from the rotational states of the molecule. The value of $|\bra{\psi}\ket{1}|^2$ describes the occupation probability of a particle on site $\ket{1}$, and $\Omega_{12}$ 
describes the tunnelling rate between $\ket{1}$ and $\ket{2}$. The detuning $\Delta_{12}$ introduces an on-site energy to the state $\ket{2}$ and so we understand our overall system as a tilted lattice. By scanning the probe microwave field we are then able to view the "many-body" spectrum of the simulation, revealing both the energy of a valence and conduction band and their populations. Because we only coupled two states together the two ``bands'' we observe are individual states. It is expected that as the number of states coupled together is increased the size of these bands should grow.

\section{Exploration of Higher Rotational States}
\begin{figure}[t]
	\centering
	\includegraphics[width=0.45\textwidth]{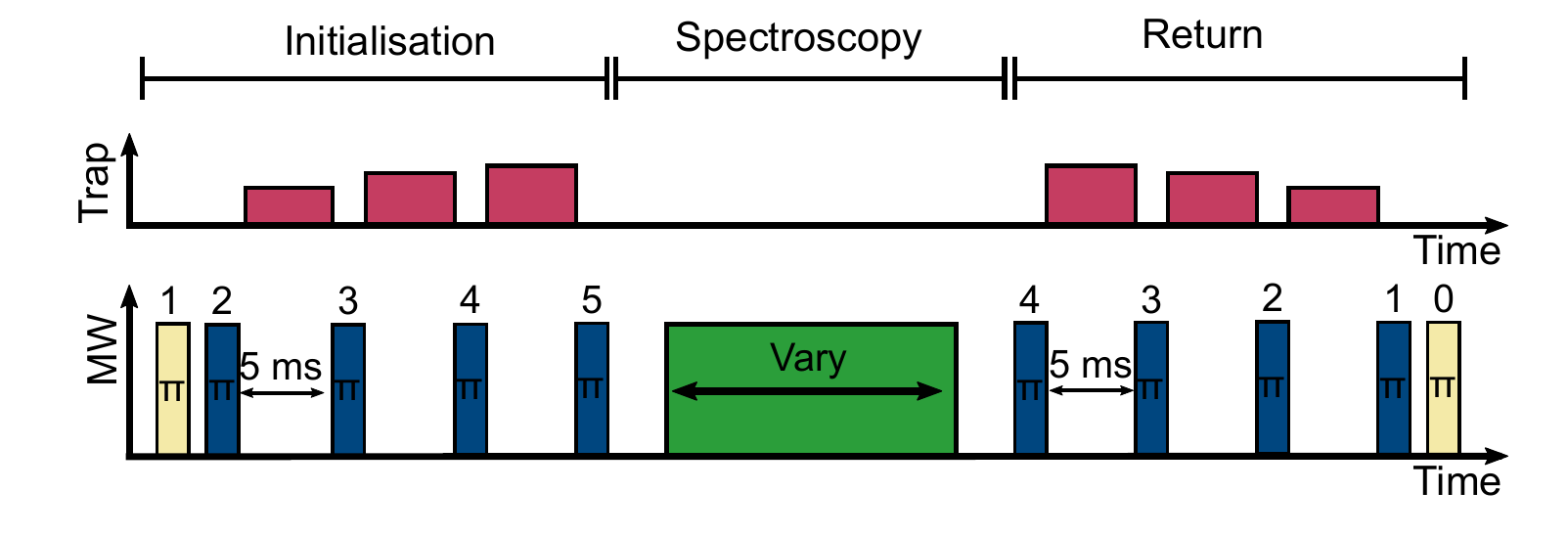}
	\caption{The sequence used to perform spectroscopy of $(N,M_F)_i=(6,11)_0$. The sequence is broken into three segments. The ``initialisation'' segment involves the transfer from $(0,5)_0$ to $(5,10)_0$ using a series of $\pi$-pulses. Each microwave $\pi$-pulse is labelled by the target rotational state. Pulses produced by microwave source A are shown in yellow, source B in blue, and source C in green. The microwave pulse generated by source C during the second ``spectroscopy'' segment has a variable frequency and/or duration. The final ``return'' segment transfers any molecules remaining in $(5,10)_0$ back to $(0,5)_0$ for detection. Between microwave pulses generated from the same source, the optical trap is switched on for 5~ms. This prevents the molecules from being lost during the time needed to switch the microwave frequency between pulses.
	}
	\label{fig:Pulses}
\end{figure}

\begin{figure}[ht!]
	\centering
	\includegraphics[width=0.4\textwidth]{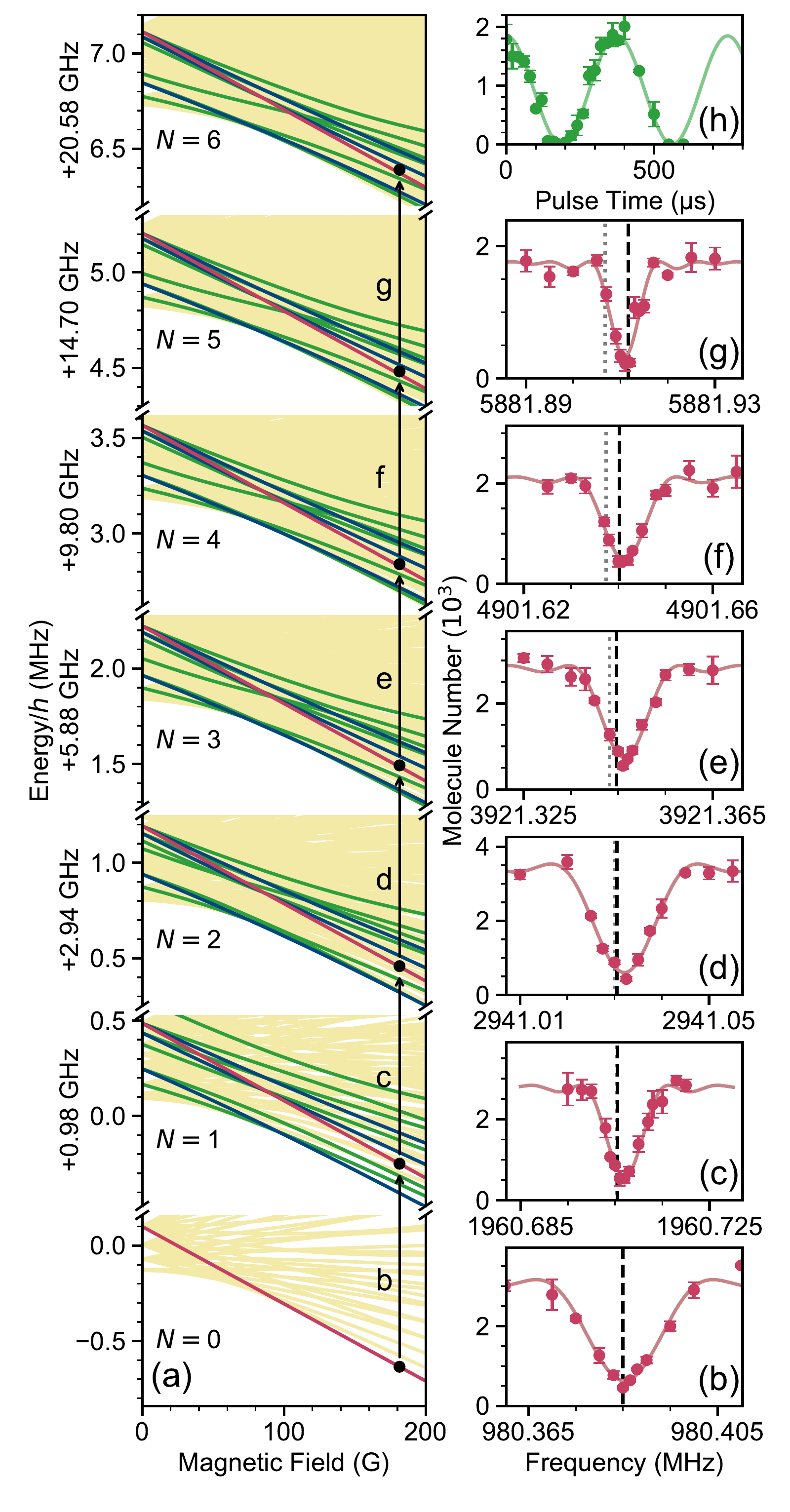}
	\caption{Spectroscopy of higher rotational states. (a) Breit-Rabi diagram for the hyperfine and Zeeman structure of the rotational states $N\le6$. The red lines indicate the sub-levels for which $M_F$ has a maximal value. Blue (green) lines indicate states where $M_F$ is one (two) units less than the maximum value for a given $N$. It is possible to couple to these states if either the microwave frequency or polarisation is not controlled.
		The points indicate the measured energies from the spectroscopy. (b) Spectroscopy of the spin-stretched transition $(N=0,M_F=5)_{i=0}\rightarrow(1,6)_0$. (c) The $(1,6)_0\rightarrow(2,7)_0$ transition. (d) The $(2,7)\rightarrow(3,8)_0$ transition. (e) The $(3,8)_0\rightarrow(4,9)_0$ transition. (f) The $(4,9)_0\rightarrow(5,10)_0$ transition. (g) The $(5,10)_0\rightarrow(6,11)_0$ transition. 
		Each spectroscopic measurement is fitted to a sinc function, where the resonant frequency and Rabi frequency are free parameters. The gray vertical dotted line indicates the theoretical transition frequency extracted from (a) using the parameters of Gregory~\textit{et al.}, the black dashed lines indicate the transition frequency extracted using our revised value of $D_v$. The numerical values of the theoretical and experimentally measured transition frequencies are reported in Table~\ref{tab:freq}. (h) Rabi oscillations on the $(5,10)_0\rightarrow(6,11)_0$ transition. In (b-h) the points are the average of three repeats, error bars show the standard error and the solid lines show fits to these results.
	}
	\label{fig:ladder}
\end{figure}

The large number of rotational states available in molecules is highly attractive for the implementation of quantum simulation schemes which utilise synthetic dimensions\cite{Boada2012,Celi2014}. Synthetic dimensions realised in atomic systems have so far been restricted to at most three states\cite{Stuhl2015,Mancini2015,Livi2016,Kolkowitz2017}, where the limit is set by the number of atomic hyperfine states available. For ultracold molecules however, it is predicted that a synthetic dimension consisting of hundreds of rotational states is feasible\cite{Sundar2018}. Furthermore, Sundar~\textit{et~al.} have shown that combining real and synthetic dimensions in systems of ultracold polar molecules can lead to the appearance of quantum strings or membranes~\cite{Sundar2018, Sundar2019}. 

Extension of the synthetic dimension beyond the two sites we have achieved requires the simultaneous coupling of greater numbers of rotational states. In this section, we therefore report spectroscopy and coherent population transfer up to the $N=6$ rotationally excited state. We choose to focus on the transitions to spin-stretched states, where $M_N$, $m_\mathrm{Rb}$ and $m_\mathrm{Cs}$ all take their maximum value. The primary reason for choosing these states is that they are insensitive to fluctuations in the magnetic field; the magnetic moments differ only by the contribution from $M_N$. The differential moment between states is therefore $g_r\mu_N\approx h\times5~\mathrm{Hz\,G^{-1}}$. We also note that if we were able to generate suitably polarised microwaves at the position of the molecules, using the spin-stretched states would allow off-resonant excitations to be completely negated. 

To access a given rotational state $N'$ we first perform a series of coherent $\pi$-pulses to $N'-1$, as illustrated in \figref{fig:Pulses} for the case of transfer to $N=5$. Each microwave transfer is performed in free-space to remove differential AC Stark shifts which would vary spatially across the cloud. Two separate microwave sources are used to drive the transfers $N=0\leftrightarrow1$ (source A, shown as yellow in \figref{fig:Pulses}) and $1\leftrightarrow2$ (source B, blue in \figref{fig:Pulses}). This allows us to transfer to $N=2$ using two sequential $\pi$-pulses with no hold time in between. The microwaves for transfer to all states $N\geq2$ are generated by source B. For this reason, after transfer to $N=2$, we recapture the molecules in the optical dipole trap for 5~ms to allow for the output of the signal generator to switch to the next frequency required. This sequence of microwave transfer and trap recapture is then repeated until the molecules occupy the desired rotational state. For each recapture, we tune the intensity of the trap to maintain the same trap parameters, compensating for the difference in polarisability between the different rotational states~\cite{Gregory2017,Blackmore2020}. For a typical transfer, the dipole trap is switched off for $<500$\,$\upmu$s, which is short enough that we do not observe significant molecule losses associated with the switching; the trap frequencies in the trap are $(\omega_x, \omega_y, \omega_z)/(2\pi)=(28, 113, 111)$\,Hz.

Once molecules occupy the state $N'-1$, we perform spectroscopy of $N'$ using a microwave pulse of variable frequency controlled by a third microwave source (source C, shown as green in \figref{fig:Pulses}). The duration and intensity of the spectroscopy pulse is set to be both less than a $\pi$-pulse when close to resonance and long enough that the Fourier width is significantly less than the approximately 50~kHz spacing between adjacent transitions. As we can only image molecules in $(0,5)_0$, following the spectroscopy pulse we must reverse the series of $\pi$-pulses to return the molecules back to $N=0$ prior to imaging, as shown in~\figref{fig:Pulses}.

We perform spectroscopy using this method for rotational states up to $N=6$, as shown in~\figref{fig:ladder}(b-g). In addition, \figref{fig:ladder}(h) shows an example of Rabi oscillations on the highest rotational transition reached, between $(5,10)_0$ and $(6,11)_0$, observed by setting the microwaves on resonance and varying the duration of the microwave pulse. As there is no decay from any of the rotationally excited states all of our spectroscopy is Fourier-transform limited. Therefore to extract centre frequencies we fit a sinc function with a width constrained by the length of the square microwave pulse to our data. We compare our spectroscopic measurements of the transition frequencies to those predicted by our model \eqref{eq:Hamiltonian}, using the hyperfine coefficients given in Gregory~\textit{et~al.}\cite{Gregory2016}, as shown by the grey dotted lines in~\figref{fig:ladder}(b-g).
The predictions using these constants appear to be accurate to less than 5~kHz for each transition frequency, and always provides an underestimate ($\chi^2_\textrm{red} =200$). To estimate the error on the predicted transition frequencies we use a Monte-Carlo method. Each parameter is sampled from a distribution with a mean and standard deviation corresponding to the best-fit value and error in Gregory~\textit{et al.}\cite{Gregory2016}. For each set of parameters, we diagonalise the resulting Hamiltonian and record the eigenenergies, labelling each eigenstate by the quantum numbers $(N,M_F)_j$. After 100 iterations we compute the mean and standard deviation of the transition frequencies. This analysis indicates that the uncertainty on the theoretical predictions of the transition frequencies, due to uncertainty in the hyperfine constants, is only a few hundred hertz. We therefore conclude that there is a statistically significant disagreement between the prediction using these parameters and our experimental measurements.

\begin{table*}[ht]
	\centering
	\caption{
		The transitions investigated in \figref{fig:ladder}(b-g) and the centre frequencies extracted from the data.
		The final two columns show the transition frequency predicted by our model using the revised value of $D_v=h\times207.3(2)~\mathrm{Hz}$ and the difference between the experiment and this value. 
	}\label{tab:freq}
	\begin{tabular}{
			l D{.}{\rightarrow}{6.6} D{.}{.}{4.6} D{.}{.}{4.6} c
		}
		\hline
		&\multicolumn{1}{c}{Transition}& \multicolumn{1}{c}{Measured} & 
		\multicolumn{1}{c}{Theoretical}& \multicolumn{1}{c}{Deviation (kHz)}\\
		&(N,M_F)_i.(N',M_F')_j&\multicolumn{1}{c}{Frequency (MHz)} &
		\multicolumn{1}{c}{Frequency (MHz)}&\\
		\hline
		(b)&(0,5)_0.(1,6)_0   &  980.385\,3(4)       &
		980.385\,00(14)&0.3(4)\\
		
		(c)&(1,6)_0.(2,7)_0   & 1960.707\,0(2)       &
		1960.705\,55(18)&1.5(3)\\
		
		(d)&(2,7)_0.(3,8)_0   & 2941.032\,1(2)       &
		2941.030\,6(3)&1.5(4)\\
		
		(e)&(3,8)_0.(4,9)_0   & 3921.346\,3(2)       &
		3921.344\,6(4)&1.7(4)\\
		
		(f)&(4,9)_0.(5,10)_0  & 4901.641\,3(3)       & 
		4901.640\,3(5)&1.0(6)\\
		
		(g)&(5,10)_0.(6,11)_0 & 5881.910\,9(2)       &
		5881.911\,7(6)&-0.8(6)\\
		
		\hline
	\end{tabular}
\end{table*}

To elucidate this discrepancy, we consider the limitations of our model in more detail. The parameters that are used in the model come from a variety of sources. The parameters that make significant contributions to the $N=0\rightarrow 1$ transition frequencies were extracted by fitting the model to high-precision microwave spectroscopy~\cite{Gregory2016}. The remaining parameters are either from ab-initio calculations~\cite{Aldegunde2017} or from conventional laser spectroscopy~\cite{Fellows1999}. One of the parameters that we did not fit in our previous microwave spectroscopy work was the centrifugal distortion term because it only contributes $h\times 800~\mathrm{Hz}$ to the energy of $N=1$. However, because the centrifugal distortion energy grows as $[N(N+1)]^2$, this term is far more significant for the transitions between higher excited states. For example, it contributes approximately 180~kHz to the $N=5\rightarrow 6$ transition frequency. A small change in the value of $D_v$ can therefore account for the deviations observed for transition frequencies higher up the rotational ladder, without impacting on the interpretation of our previous $N=0\rightarrow 1$ measurements. Allowing the centrifugal distortion term to vary in our fit to the spectroscopy measurements  gives a revised value of $D_v = h\times 207.3(2)~\textrm{Hz}$ ($\chi^2_\textrm{red} = 40$). This is $h\times5.7(4)~\mathrm{Hz}$ smaller than the value reported by Fellows~\textit{et al.}~\cite{Fellows1999}. 
The remaining terms in the Hamiltonian~\eqref{eq:Hamiltonian} do not have a large enough impact on the transition frequencies when varied within their respective uncertainties. For direct comparison we tabulate the experimentally measured and updated predictions of the transition frequencies in Table~\ref{tab:freq}.

\begin{figure}[t]
	\centering
	\includegraphics[width=0.45\textwidth]{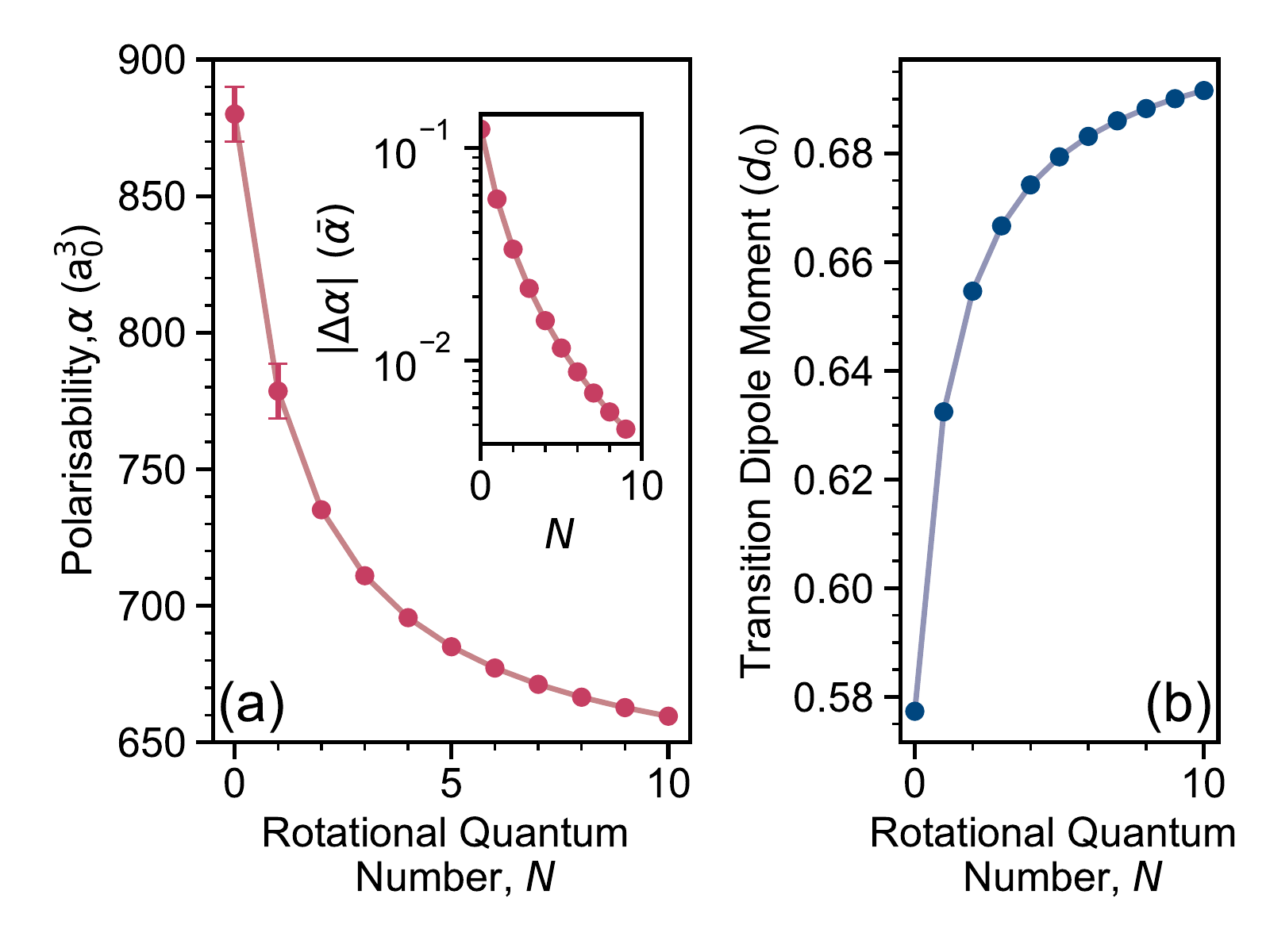}
	\caption{Key properties of the spin-stretched states $\ket{N,M_N=N}$. (a)~The polarisability for a $\lambda=1550$\,nm laser, linearly polarised along the magnetic field, as a function of $N$. The error bars at $N=0$ and $N=1$ indicate the experimental error from the values of $\alpha^{(0)}$ and $\alpha^{(2)}$ measured previously~\cite{Gregory2017}.
		Inset: The differential polarisability, $\Delta\alpha = \alpha(N)-\alpha(N+1)$, between neighbouring rotational states in units of the mean polarisability $\bar{\alpha}=[\alpha(N)+\alpha(N+1)]/2$. 
		(b)~The magnitude of the transition dipole moment $|\bra{N'=N+1,M_N'=N'}\mu^+\ket{N,M_N}|$, in units of the molecule's permanent dipole moment ($d_0 =1.23~\mathrm{D}$), as a function of the rotational quantum number. }
	\label{fig:Higher_N_acstark}
\end{figure}  

In addition to increasing the number of states available, access to higher-energy rotational states has a number of benefits for future experiments. For example, the coherence time for superpositions of different rotational states is often limited by the differential polarisability between the two states~\cite{Neyenhuis2012,Blackmore2018,Seesselberg2018_Coherence}. A common approach to eliminate this source of decoherence is to set the polarisation of the trapping laser at an angle to an applied DC electric field, such that the difference in polarisability for states with $M_N=0$ tends to zero~\cite{Neyenhuis2012,Seesselberg2018_Coherence}. However, this limits the states which can be used and may not be conducive to certain trap geometries. We have calculated the polarisability for spin-stretched rotational states up to $N=10$ for the case where the laser polarisation is parallel to the applied magnetic field; the results are shown in \figref{fig:Higher_N_acstark}(a). It can be seen that the difference in polarisability between neighbouring rotational states reduces as $N$ increases. For $N>7$, the differential polarisability between $N$ and $N+1$ is less than 1\%. This represents an order of magnitude improvement over $N=0$ and $N=1$, and we expect long coherence times to be accessible without special consideration in the design of the trap. Furthermore, the magnitude of the transition dipole moment for the spin-stretched transitions also rises asymptotically towards $d_0/\sqrt{2}$ with increasing $N$ (see \figref{fig:Higher_N_acstark}(b)). This indicates that interactions such as spin-exchange can be stronger in higher-energy rotational states, and the variation with increasing $N$ could be used to tune the rate of spin-exchange in future experiments. 

\section{Conclusions}

In summary, we have studied multi-level quantum systems constructed from the rotational and hyperfine states of ultracold molecules. We have demonstrated how two-photon Raman transfer may be used to resonantly couple hyperfine levels of the rotational ground state, while avoiding population of the first rotationally excited state. Such control may be useful for initialisation of a quantum memory. We determined a Ramsey coherence time for a superposition of the qubit states for such a quantum memory of 58(9)\,ms. We then explored the generation of coherent dressed states, using a 3-level ladder configuration of rotational states. Through Autler-Townes spectroscopy, we have demonstrated the creation and coherent population of microwave dressed states in this system, and discussed how this may be interpreted in terms of future quantum simulations in synthetic dimensions. Finally, we explore the possibilities afforded by going to higher rotational states. We have performed spectroscopy of the rotational states of RbCs up to $N=6$, and in doing so have determined a refined value of the centrifugal distortion coefficient $D_v=h\times207.3(2)$\,Hz. This work contributes to the continuing efforts to develop more advanced techniques for coherent control of the quantum states of ultracold molecules, which will be crucial for future applications in the fields of quantum computation and quantum simulation.

\section*{Conflicts of interest}
The authors declare no conflicts of interest.
\section*{Acknowledgements}
The authors would like to thank Elizabeth Bridge and Rahul Sawant for contributions to early stages of this work. We thank Ifan Hughes for suggesting the Monte-Carlo method for analysis of theoretical errors. We also acknowledge stimulating discussions with Kaden Hazzard, Lincoln Carr and Jeremy Hutson. This work was supported by U.K. Engineering and Physical Sciences Research Council (EPSRC) Grants EP/P01058X/1 and EP/P008275/1.

The data, code and analysis associated with this work are available at: \textbf{DOI:10.15128/r2xg94hp56r}. The python code for hyperfine structure calculations can be found at \textbf{DOI:10.5281/ZENODO.3755881}.


\appendix
\section{Composition of Hyperfine States}\label{App:States}
In the main body of the paper we perform experiments on the rotational and hyperfine states of RbCs at a magnetic field of 181.5~G. At this magnetic field the good quantum numbers are the rotational quantum number $N$ and the projection $M_F$ of the total angular momentum. Because these quantum numbers do not uniquely identify individual hyperfine sub-levels, we label each by $(N,M_F)_k$ where $k$ is an index, starting at $k=0$, that counts up in energy at 181.5~G for states with the same values of $N$ and $M_F$. In Table~\ref{tab:states}, we give the state composition of the relevant $(N,M_F)_k$ states in the uncoupled $\ket{N,M_N,m_\mathrm{Rb},m_\mathrm{Cs}}$ basis, where $M_N$ is the projection of $N$ onto the magnetic field axis and $m_\mathrm{Rb}$ and $m_\mathrm{Cs}$ are the projections of the \textsuperscript{87}Rb and \textsuperscript{133}Cs nuclear spins onto the magnetic field axis. The spin stretched states, $(N,M_F=N+5)_0$, are all represented by the single nuclear spin state $\ket{m_\mathrm{Rb}=3/2,m_\mathrm{Cs}=7/2}$.

\begin{table}[h]
	\centering
	\caption{The hyperfine states used for the Raman transition and Ramsey measurements presented in \figref{fig:Rabi_oscillations} and \figref{fig:Ramsey_fringes}. The calculations use the coefficients of Gregory~\textit{et al.}~\cite{Gregory2016} and coefficients are rounded to 1 part in $10^3$. Also included is the $(0,3)_0$ state which we predict would have a longer coherence time in a superposition with $(0,4)_1$.}
	\begin{tabular}{c l}
		\hline
		$(N,M_F)_k$ & Composition in the  $\ket{N,M_N,m_\mathrm{Rb},m_\mathrm{Cs}}$ basis\\
		\hline
		$(0,5)_0$& $1.0 \ket{0,0,3/2,7/2}$  \\
		$(0,4)_1$& $0.321 \ket{0,0,3/2,5/2} + 0.947 \ket{0,0,1/2,7/2}$\\
		$(1,5)_0$& $0.087 \ket{1,1,3/2,5/2} - 0.370\ket{1,1,1/2,7/2}$\\
		&$+0.925\ket{1,0,3/2,7/2}$\\
		$(0,3)_0$& $0.928 \ket{0,0,3/2,3/2} -0.365 \ket{0,0,1/2,5/2}$\\
		&$+0.074\ket{0,0,-1/2,7/2}$\\
		\hline
	\end{tabular}
	
	\label{tab:states}
\end{table}

\bibliography{MultiPhoton} 

\end{document}